\documentclass[aps, prb, preprint, superscriptaddress, showpacs]{revtex4-1}
\usepackage{graphicx}
\usepackage{natbib}

\begin{document}

\title{Role of vacancies in the magnetic and electronic properties of SiC nanoribbons: an ab initio study}
\author{Juliana M. Morbec}
\email{jmmorbec@gmail.com}
\affiliation{Instituto de Ci\^encias Exatas, Universidade Federal de Alfenas, 37130-000,
Alfenas, MG, Brazil}
\author{Gul Rahman}
\email{gulrahman@qau.edu.pk}
\affiliation{Department of Physics, Quaid-i-Azam University, Islamabad, 45320, Pakistan} 

\date{\today}

\begin{abstract}
Using {\it ab-initio} calculations based on density functional theory, we investigate the effects of vacancies on the electronic and magnetic properties of
zigzag SiC nanoribbons (Z-SiCNR). 
Single ($V_{\rm C}$ and $V_{\rm Si}$) and double ($V_{\rm Si}V_{\rm Si}$ and  $V_{\rm Si}V_{\rm C}$) vacancies 
are observed to induce magnetism in Z-SiCNRs. 
The presence of a single $V_{\rm Si}$ does not affect the half-metallic behavior of pristine Z-SiCNRs; 
however, a single $V_{\rm C}$ leads to a transition from half-metallic to metallic behavior in Z-SiCNRs due to the edge Si $p$ orbitals 
and the atoms surrounding the vacancy. 
The interactions of vacancies with foreign impurity atoms (B and N) are also investigated and it is observed that $V_{\rm Si}N_{\rm C}$ does not only suppress 
the oscillatory type magnetism of $V_{\rm Si}V_{\rm C}$, but also retains the half-metallic character of the pristine Z-SiCNRs. 
The defect formation energies of vacancies can be reduced by substitutional B and N atoms. 
We believe that ferromagnetism is expected if Z-SiCNR are grown under suitable conditions.

\end{abstract}


% 73.22.-f: Electronic structure of nanoscale materials 
% 71.55.-i: Impurities - electronic structure
% 75.75.-c: Magnetic properties of nanostructures
% 71.15.Mb: Density-functional theory - condensed matter
% 71.15.Nc: Total energy calculations (condensed matter)

\maketitle

\section{INTRODUCTION} \label{intro}

Silicon carbide (SiC) is an attractive material for numerous technological applications, mainly in harsh environments. 
Bulk SiC is known to possess outstanding properties 
(such as high thermal conductivity, high breakdown electric field, high electronic mobility, excellent chemical and physical stability, 
good radiation resistance, and wide band gap)\cite{Melinon_NatMaterial, Pierre2002_SurfSciRep} 
which make it a suitable semiconductor for high-power, high-temperature, and high-frequency devices. 
Furthermore, SiC nanowires and nanotubes, which have already been synthesized,\cite{XHSun2002, ZPan2000} 
exhibit excellent characteristics and are good candidates for applications ranging
from hydrogen storage media\cite{doi:10.1021/nl0603911} and gas sensors\cite{1674-4926-30-11-114010} 
to optical\cite{PhysRevB.84.085404} and field-emission devices.\cite{Kim2008}

In the last few years, the successful synthesis of SiC nanotubes\cite{XHSun2002} 
and, recently, the theoretical prediction of the stability of two-dimensional SiC monolayer with 
honeycomb structure\cite{PhysRevB.80.155453, PhysRevB.81.075433}  
have stimulated increasing interest in SiC nanosheets and nanoribbons (NRs). 
According to recent theoretical studies, 
SiC nanosheets as well as armchair SiCNRs (A-SiCNRs) behave as nonmagnetic wide band gap 
semiconductors,\cite{PhysRevB.80.155453, PhysRevB.81.075433, sun:174114}   
whereas zigzag SiCNRs (Z-SiCNRs) are magnetic (with very small magnetic moments)\cite{sun:174114, LouLee_12637} 
and can present metallic or semiconducting character,  
depending on the width of the nanoribbon\cite{sun:174114, Zang_2010_3259, LouLee_12637} 
(half-metallicity was predicted for Z-SiCNRs narrower than 4 nm,\cite{sun:174114} which makes these NRs 
promising candidates for spintronic applications). 
However, {\it ab initio} investigations have recently shown that these characteristics can be modified by the presence of some  
impurities and defects. 
For example, it has been observed that 
(i) substitutional B, N, As, and P impurities 
 induce magnetism in SiC sheets,\cite{PhysRevB.81.075433} 
(ii) half-metallic Z-SiCNRs become metallic when doped with N atoms,\cite{Costa_Morbec_2011} 
and (iii) B (N) substituting an edge Si (C) atom 
transforms semiconducting Z-SiCNRs into half-metallic systems.\cite{Lou91_2012} 
In addition, Si vacancy has been shown to induce magnetism in nonmagnetic SiC sheets\cite{PhysRevB.81.075433, He20102451} and A-SiCNRs.\cite{PhysRevB.81.075433}

Intrinsic defects, especially vacancies, have been suggested to be related to the origin of magnetism in SiC structures: 
besides the aforementioned studies of vacancies in SiC sheets and A-SiCNRs (Refs.~\onlinecite{PhysRevB.81.075433, He20102451}), 
there are experimental evidences that defects dominated by Si+C divacancies 
can induce room-temperature ferromagnetism in diamagnetic SiC crystals.\cite{PhysRevLett.106.087205, li:222508} 
Despite the great potential of Z-SiCNRs for spintronics, the role of intrinsic defects (vacancies) and 
their interactions with foreign impurities in these NRs 
have not, to our knowledge, been investigated so far. In this work we investigated,  
by means of {\it ab initio} calculations, the effects of vacancies on the electronic and magnetic properties of Z-SiCNRs, 
and the interactions of vacancies with foreign impurity atoms. 
Our extensive results indicate that the presence of one C ($V_{\rm C}$) or Si ($V_{\rm Si}$) vacancy per supercell, 
as well as two vacancies ($V_{\rm Si}V_{\rm Si}$ or $V_{\rm Si}V_{\rm C}$), can induce large magnetic moments in Z-SiCNRs. 
The defect formation energy of $V_{\rm Si}$  is decreased  when the interactions with B and N impurities are considered.

\section{COMPUTATIONAL DETAILS}

\begin{figure}[!h]
\centering
\includegraphics[scale=0.35]{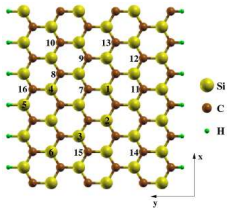}
\caption{\label{supercell} 
(Color online) Structural model of 6Z-SiCNR. 
Labels $1,2,...,16$ indicate the positions of the vacancies or impurities.}
\end{figure}

Spin-polarized calculations were performed within the framework of the density functional theory (DFT),\cite{HohenbergKohn}  
as implemented in the SIESTA code,\cite{Siesta2002} employing the 
generalized gradient approximation of Perdew, Burke, and Ernzerhof\cite{GGAPBE} (GGA-PBE) for the exchange-correlation functional and 
norm-conserving Troullier-Martins pseudopotentials\cite{TroullierMartins1991} to describe the electron-ion interactions. 
We used an energy cutoff of 200 Ry for the real-space mesh and a double-zeta basis set with polarization functions for all atoms.  
We considered Z-SiCNR with width W = 6 (6Z-SiCNR), as depicted in Fig.~\ref{supercell}.  
The 6Z-SiCNR was simulated within the supercell approach with 84 atoms in the unit cell (including H atoms to passivate 
the edge dangling bonds) and vacuum regions of $17$ \AA \ and $20$ \AA \ 
along the y and z directions, respectively. 
{Additional test calculations were also carried out for large supercells (12Z-SiCNR) containing 156 atoms to see the convergence of our numerical results.} 
All atomic positions were fully relaxed until the forces on each atom were smaller than 0.05~eV/\AA. 
The Brillouin zone was sampled using Monkhorst-Pack $k$-point meshes of $11 \times 1 \times 1$ and $121 \times 1 \times 1$ 
for the total-energy and electronic-structure calculations, respectively. 
Using the Mulliken population analysis, the local magnetic moment was calculated by $m=\rho^{\uparrow}-\rho^{\downarrow}$, 
where $\rho^{\uparrow}$ ($\rho^{\downarrow}$) represents the spin-up (-down) valence electrons of an atom. 

In this work, we investigated the presence of single (one $V_{\rm C}$ or $V_{\rm Si}$ per supercell) and double 
($V_{\rm Si}V_{\rm Si}$ and $V_{\rm Si}V_{\rm C}$) vacancies in 6Z-SiCNR. 
The energetic stability of these systems was examined by calculating their defect formation energies $E_f$. 
We used 
\begin{equation}
E_f=E[\mbox{$V_{\rm X}$/6Z-SiCNR}] - E[\mbox{6Z-SiCNR}] + \mu_{\rm X},
\label{ef_single}
\end{equation} 
for single vacancy (X = Si or C), and
\begin{equation}
E_f=\frac{1}{n}\left\{E[\mbox{$V_{\rm Si}V_{\rm X}$/6Z-SiCNR}] - E[\mbox{6Z-SiCNR}] +\mu_{\rm Si}+\mu_{\rm X}\right\}
\label{ef_double}
\end{equation}
for double $V_{\rm Si}V_{\rm X}$ ($n=2$ if  X = Si, and $n=1$ if X = C). 
Here, $E[\mbox{6Z-SiCNR}]$ is the total energy of pristine 6Z-SiCNR, and  
$E[\mbox{$V_{\rm X}$/6Z-SiCNR}]$ and $E[\mbox{$V_{\rm Si}V_{\rm X}$/6Z-SiCNR}]$ are the total energies of 
6Z-SiCNR with single $V_{\rm X}$ and double $V_{\rm Si}V_{\rm X}$, respectively. 
The chemical potential of the atomic specie $X$ is $\mu_{\rm X}$. 
{In equilibrium conditions, 
\begin{equation}
\mu_{\rm Si}+\mu_{\rm C}=\mu_{\rm SiC}^{\rm NR}=\mu_{\rm Si}^{\rm bulk} + \mu_{\rm C}^{\rm bulk} -\Delta H_f.
\end{equation} 
$\mu_{\rm Si}$ and $\mu_{\rm C}$ characterize the  growth conditions: 
in the Si-rich limit, the system is assumed to be in thermodynamic equilibrium with the bulk Si and $\mu_{\rm Si}=\mu_{\rm Si}^{\rm bulk}$ 
(in this case, $\mu_{\rm C}=\mu_{\rm SiC}^{\rm NR}-\mu_{\rm Si}^{\rm bulk}$); in the C-rich limit, $\mu_{\rm C}=\mu_{\rm C}^{\rm bulk}$ 
and $\mu_{\rm Si}=\mu_{\rm SiC}^{\rm NR}-\mu_{\rm C}^{\rm bulk}$. 
We considered diamond structures in the calculation of $\mu_{\rm C}^{\rm bulk}$ and $\mu_{\rm Si}^{\rm bulk}$. 
$\Delta H_f$ is the heat of formation of the SiCNR. We found $\Delta H_f=-1.214$~eV, which compares well with the heat of formation of 
SiC nanotubes (-1.2 and -1.3~eV for (6,6) and (8,0) SiC nanotubes, respectively).\cite{PhysRevB.73.245415}  
To check the quality of the pseudopotentials and computational parameters which are used in the present calculations, 
we also calculated the heat of formation of bulk 3C-SiC; we found $\Delta H_f = 0.679$~eV, 
which is in good agreement with the experimental value of 0.72~eV.\cite{PhysRevB.55.10561} }

\section{RESULTS AND DISCUSSION} \label{results}

\begin{figure}[!b]
\centering
\includegraphics[scale=0.34]{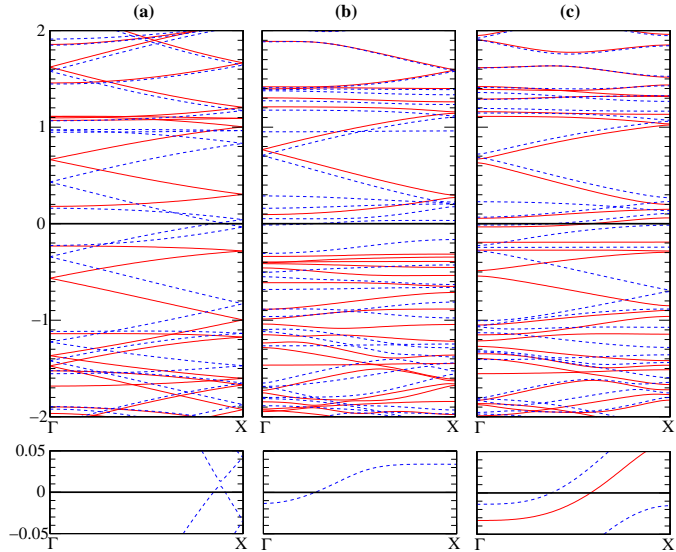}
\caption{\label{bands} 
(Color online) Electronic band structure of (a) pristine 6Z-SiCNR, (b) $V_{\rm Si}$/6Z-SiCNR, and (c) $V_{\rm C}$/6Z-SiCNR.
The band structure of each system is presented in two ranges: 
$|E - E_F| \leq  2.0$~eV (top panel) and $|E - E_F| \leq 0.05$~eV (bottom panel). 
Solid red and dashed blue lines indicate spin-up and spin-down bands, respectively. The Fermi level is set to zero and 
indicated by the solid black line.}
\end{figure}

\begin{figure}[!t]
\centering
\includegraphics[scale=0.35]{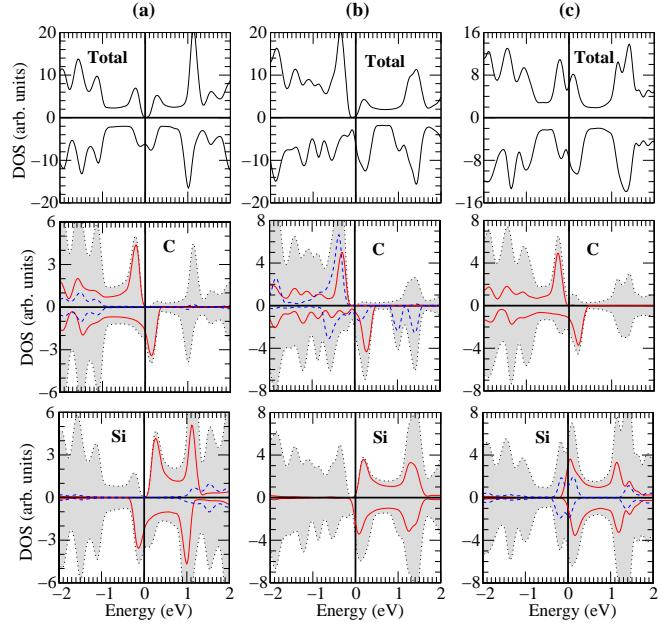}
\caption{\label{density_states} 
(Color online) Density of states (DOS)  of (a) pristine 6Z-SiCNR, (b) $V_{\rm Si}$/6Z-SiCNR, and (c) $V_{\rm C}$/6Z-SiCNR. 
In the middle [bottom] panels the shaded regions indicate the DOS of the C [Si] atoms; solid red lines show $p$ orbitals of the edge C [Si] atoms; and 
dashed blue lines show $p$ orbitals of the three C [Si] atoms around Si$^{1}$ [C$^7$] in (a) and $V_{\rm Si}^1$ [$V_{\rm C}^7$] in (b) [(c)]. 
The Fermi level is set to zero and indicated by the solid black line. 
We considered a Gaussian broadening of 0.1~eV for the DOS diagrams. }
\end{figure}

\begin{figure*}[!t]
\centering
\includegraphics[scale=0.33]{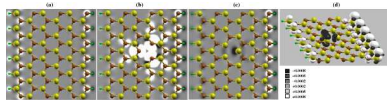}
\caption{\label{spin_density} 
(Color online) Contour plot of spin-density distribution of (a) pristine 6Z-SiCNR, 
(b) $V_{\rm Si}$/6Z-SiCNR, and (c) $V_{\rm C}$/6Z-SiCNR. 
For comparison purposes the spin-density isosurface of $V_{\rm C}$/6Z-SiCNR system is also shown in (d) at  $8.0\times 10^{-4}$~e/Bohr$^{3}$.}
\end{figure*}

First, we examined the electronic and magnetic properties of pristine 6Z-SiCNR. 
The electronic band structure [Fig.~\ref{bands}(a)] and the density of states [Fig.~\ref{density_states}(a)] show that 6Z-SiCNR 
exhibits half-metallic behavior: the spin-up channel is semiconducting (with a direct band gap of about 0.41 eV 
at the $\Gamma$ point) whereas the spin-down channel is metallic. 
As can be seen in Fig.~\ref{density_states}(a), the partially occupied spin-down electronic state 
is mainly composed of $p$ orbitals of the edge C and Si atoms.
The 6Z-SiCNR also has a very small magnetic moment, about $0.03\mu_B$ per supercell. 
Our results are in agreement with a previous theoretical work\cite{sun:174114} which reports magnetic moment of 
0.023$\mu_B$ per cell in 6Z-SiCNR, and predicts half-metallicity in Z-SiCNRs narrower than 4~nm.
The spin-density distribution of pristine 6Z-SiCNR [Fig.~\ref{spin_density}(a)] shows that the magnetic moments are mainly 
localized on the edge atoms, and their orientations are parallel at each edge and antiparallel between the two edge atoms. 
Each edge C (Si) atom has a local magnetic moment of about $0.16\mu_B$ ($-0.17\mu_B$), while the total magnetic moment of the other C (Si) atoms is 
about  $0.29\mu_B$ ($-0.20\mu_B$). The strong contribution of the edge C and Si atoms to the magnetization of 6Z-SiCNR 
can also be seen in Fig.~\ref{density_states}(a): the spin-up $p$ states of the edge C atoms are completely occupied whereas the spin-down states 
are partially occupied; for the edge Si atoms, we note that their spin-down $p$  states have a small occupation whereas 
their spin-up states are empty.

\subsection{Single vacancy}

{We have investigated the effects of vacancies on the electronic and magnetic  properties of 6Z-SiCNR. } 
We considered initially one $V_{\rm Si}$ or $V_{\rm C}$ per supercell, 
which corresponds to a defect concentration of 2.78\%. 
$V_{\rm Si}$ and $V_{\rm C}$ were created at the center of the nanoribbon, by removing Si and C atoms labeled 1 and 7 [see Fig.~\ref{supercell}], respectively.  
The presence of a single $V_{\rm Si}$ and $V_{\rm C}$ induces magnetic moments of 4.19 and $1.19\mu_B$ per supercell, respectively. 
We found that one $V_{\rm Si}$ per supercell leads to an increase, from 0.03 to $4.19\mu_B$, in the magnetic moment of 6Z-SiCNR, which  
is similar to that reported for SiC sheets and A-SiCNRs in previous {\it ab initio} studies where a single $V_{\rm Si}$ was found to  
induce a large magnetic moment of $4.0\mu_B$ in nonmagnetic SiC sheets\cite{PhysRevB.81.075433, He20102451} and A-SiCNRs.\cite{PhysRevB.81.075433} 
On the other hand, $V_{\rm C}$ appears to play different roles in the magnetism of Z-SiCNRs, A-SiCNRs and SiC sheets: 
while we found a magnetic moment of $1.19\mu_B$ induced by a single $V_{\rm C}$ in 6Z-SiCNRs,
recent theoretical works have reported that the presence of a single $V_{\rm C}$ does not give rise to any magnetic 
moment in SiC sheets\cite{PhysRevB.81.075433, He20102451} and A-SiCNRs.\cite{PhysRevB.81.075433} 
This suggests that 6Z-SiCNR, where magnetism can be tuned either by Si or C vacancies, can be a promising candidate for spintronics. 
{In the following paragraphs we will discuss the origin of the magnetism in the V${\rm_C}$/6Z-SiCNR system, 
showing that it is almost entirely due to the magnetization of the edge C atoms.} 
{As we have shown that V${\rm_C}$/6Z-SiCNR system is magnetic. 
To be more confident about the magnetism of V${\rm_C}$/6Z-SiCNR, we also carried out calculations using a large supercell (156 atoms) and created a 
carbon vacancy and relaxed all the atoms. Interestingly, we found that this system also leads to magnetism in SiCNR and the magnetism was mainly 
originated from the edge C atoms. Therefore, we confirm that the magnetism of V${\rm_C}$/6Z-SiCNR  is not due to structural relaxation, 
but due to the C edge atoms.}  

Figure~\ref{spin_density}(b) presents the distribution of spin density in 6Z-SiCNR with single $V_{\rm Si}$ ($V_{\rm Si}$/6Z-SiCNR). 
We note that the spin density is mostly localized around the vacancy and on the edge C atoms:  
the local magnetic moment of the each edge C atom varies from 0.233 to $0.287\mu_B$, while 
the three C atoms surrounding the $V_{\rm Si}$ have magnetic moments of 0.435, 0.865 and $0.865\mu_B$. 
This can also be seen in Fig.~\ref{density_states}(b), which shows that the spin-up $p$ states of the edge C atoms and of the 
C atoms around the $V_{\rm Si}$ are completely occupied whereas their spin-down states are partially occupied. The presence of a single $V_{\rm Si}$ 
destroys the complete occupation of $p$ orbitals of the C atoms surrounding the vacancy, giving rise to empty spin-down states above the Fermi level.
The local magnetic moment of each edge Si atom is about $-0.04\mu_B$,  which suggests that the C atoms and the Si atoms couple antiferromagnetically. 
Such types of magnetic coupling are also observed in the other defect driven magnetic systems.~\cite{Stefano,Zhao,Tatsuya,GR}
The smaller magnetization of the edge Si atoms in $V_{\rm Si}$/6Z-SiCNR, compared to that observed in pristine 6Z-SiCNR (where we found 
magnetic moments of about $-0.17\mu_B$ for each edge Si atom), can also be seen in Fig.~\ref{density_states}; by comparing Figs.~\ref{density_states}(a) and (b) 
we note that the occupation of spin-down $p$ states of the edge Si atoms is larger in pristine 6Z-SiCNR [Fig.~\ref{density_states}(a)] 
than in $V_{\rm Si}$/6Z-SiCNR [Fig.~\ref{density_states}(b)]. 
Such changes in the occupation of spin-down $p$ states also decrease the exchange splitting of the edge Si atoms which result in small magnetic moments. 
On the other hand, the exchange splitting is large in the edge C atoms. 

In the $V_{\rm C}$/6Z-SiCNR system we found that the magnetic moment comes mainly from the edge C 
atoms [see Figs.~\ref{density_states}(c), \ref{spin_density}(c), and \ref{spin_density}(d)]. 
Each edge C atom has magnetic moment of about $0.20\mu_B$, whereas 
the Si atoms surrounding the $V_{\rm C}$ have magnetic moments of -0.04, -0.07 and $-0.07\mu_B$. 
It is interesting to see that both the Si and C atoms are isoelectronic where one could expect the same magnetism in both Si and C vacancies. 
However, we observed a quite different nature of magnetism, e.g., atoms surrounding $V_{\rm Si}$ ($V_{\rm C}$) have large (small) local positive (negative) 
magnetic moments. This behavior can be understood due to different nature of wave functions of the $p$ orbitals. 
Usually, cation/anion vacancies induce spin positive polarization on the surrounding atoms, but $V_{\rm C}$/6Z-SiCNR shows 
the spin density of the Si atoms is larger in the spin down states. 

The electronic band structure [Fig.~\ref{bands}(b)] and the density of states [Fig.~\ref{density_states}(b)] 
of $V_{\rm Si}$/6Z-SiCNR show that the presence of a single $V_{\rm Si}$ does not affect the half-metallic character of pristine 6Z-SiCNR; 
in this system, the spin-down channel is metallic, whereas the spin-up channel is semiconducting, 
with indirect band gap of 0.41~eV.  
In contrast, we found that a single $V_{\rm C}$ transforms the half-metallic 6Z-SiCNR into a metallic system 
[see Figs.~\ref{bands}(c) and \ref{density_states}(c)]. 
Note that the half-metallicity is mainly destroyed by the $p$ orbitals of the edge atoms and the atoms surrounding the vacancy.
Figure~\ref{density_states}(c) shows that the partially occupied spin-up electronic state 
consists mainly of $p$ orbitals of the edge Si atoms and of the Si atoms surrounding the vacancy, with no contribution from $p$ orbitals of the 
edge C atoms. 
This shows that the electronic structure can easily be tuned either by creating holes (simply based on electron counting), i.e., $V_{\rm C}$, 
or by electron doping, i.e., 
N-doping.\cite{Costa_Morbec_2011} We attribute this electronic phase transition to hole doping.\cite{footnote} 
The electronic phase transition is mainly due to  
Si-$p$ orbitals which are delocalized as compared with the C-$p$ orbitals. 
The total densities of states in Fig.~\ref{density_states}(a--c) show that the population of density of states 
(within an energy range of   $\pm0.5$ eV)  in spin-up states of electrons 
increases as we move from pristine to $V_{\rm C}$ system. This increment is mainly caused by the edge Si atoms and Si atoms surrounding the vacancy. 

The energetic stability of single $V_{\rm Si}$ and $V_{\rm C}$ in 6Z-SiCNR was examined by comparing $E_f$ of 
the $V_{\rm Si}$/6Z-SiCNR and $V_{\rm C}$/6Z-SiCNR systems.
Our results for $E_f$, obtained using Eq.~(\ref{ef_single}), are listed in Table~\ref{table1}. 
As can be seen, the occurrence of a single $V_{\rm C}$ in Z-SiCNRs is energetically favored over single $V_{\rm Si}$. 
This behavior is similar to that observed in bulk SiC,\cite{PhysRevB.59.15166} SiC sheet\cite{He20102451} and SiC nanotubes,\cite{PhysRevB.74.155425} where 
$V_{\rm C}$ has also been reported to be more stable than $V_{\rm Si}$.

\subsection{Double vacancies}

In order to study the interactions of $V_{\rm Si}$ and $V_{\rm C}$, we also investigated the presence of two vacancies per supercell. 
Since $V_{\rm Si}$ appears to play a more important role on the magnetism of Z-SiCNR than $V_{\rm C}$ we considered 
configurations with (i) two Si vacancies, $V_{\rm Si}^1V_{\rm Si}^i$, and (ii) Si + C vacancies, $V_{\rm Si}^1V_{\rm C}^i$. 
The index $i$ in $V_{\rm X}^i$ indicates the position of the $X$ vacancy [see Fig.~\ref{supercell}]. 

The formation energies and the total magnetic moments per supercell 
for $V_{\rm Si}^1V_{\rm Si}^i$ and $V_{\rm Si}^1V_{\rm C}^i$ in 6Z-SiCNR are presented in Table \ref{table1} 
(the defect formation energies were calculated using Eq.~(\ref{ef_double})). 
These results show that the presence of double Si vacancies can give rise to large magnetic moments in Z-SiCNRs  
(for example, we found magnetic moments of about $3.9\mu_B$ in \mbox{$V_{\rm Si}^1V_{\rm Si}^3$/6Z-SiCNR} 
and \mbox{$V_{\rm Si}^1V_{\rm Si}^4$/6Z-SiCNR}). 
However, the formation of these magnetic \mbox{$V_{\rm Si}^1V_{\rm Si}^i$/6Z-SiCNR} systems is quite unlikely, 
since they have very high formation energies [see  Table~\ref{table1}]. 
It is also noticeable the $V_{\rm Si}-V_{\rm Si}$ interactions reduce the magnetic moments as compared with the single $V_{\rm Si}$ system.
The formation of $V_{\rm Si}^1V_{\rm C}^i$, which also induce magnetic moments in 6Z-SiCNRs,   
is energetically favorable when compared with a single $V_{\rm Si}$ and double $V_{\rm Si}^1V_{\rm Si}^i$. 
{For instance, $V_{\rm Si}^1V_{\rm C}^8$, $V_{\rm Si}^1V_{\rm C}^{12}$, and 
$V_{\rm Si}^1V_{\rm C}^{13}$ have formation energies smaller than 4.7~eV, and induce magnetic moments of 1.34, -0.93 and -0.57$\mu_B$ in 6Z-SiCNR, respectively 
[see Table~\ref{table1}].}
It is worth mentioning here that recent experimental works have shown that defects created in 6H-SiC bulk by neutron or ion irradiations 
are mainly formed by $V_{\rm Si}V_{\rm C}$ divacancies and induce room-temperature magnetism in diamagnetic SiC crystals.\cite{PhysRevLett.106.087205, li:222508} 
Such experimental results illustrate that $V_{\rm Si}V_{\rm C}$ are the stable intrinsic defects in 6H-SiC, and we also found that $V_{\rm Si}V_{\rm C}$ 
types defects are easy to be formed in NR as compared with $V_{\rm Si}$. 
Table~\ref{table1} clearly shows that $V_{\rm Si}V_{\rm C}$ divacancies have oscillatory type magnetism where the magnetic moments oscillate and 
strongly depend on the location of the C vacancy. This rise and fall of magnetic moments in $V_{\rm Si}V_{\rm C}$ divacancies system is  
similar to bulk SiC crystal.\cite{li:222508} Comparing the defects, we note that this rise and fall of magnetic moments is 
suppressed by filling the C vacant site with N, which will be discussed in the following paragraphs. 

\begin{table}[!t]
\caption{\label{table1} The calculated formation energies, in Si-rich and C-rich conditions, and total magnetic moments per supercell ($M$)  
for single ($V_{\rm Si}$ and $V_{\rm C}$) and double ($V_{\rm Si}^1V_{\rm Si}^i$ and $V_{\rm Si}^1V_{\rm C}^i$) vacancies in 6Z-SiCNR. 
The index $i$ in $V_{\rm X}^i$ indicates the position of the vacancy [see Fig.~\ref{supercell}]. 
Single $V_{\rm Si}$ ($V_{\rm C}$) is localized on site 1 (7).}
\begin{ruledtabular}
\begin{tabular}{lccc}
              &  & \multicolumn{2}{c}{Formation energies (eV)} \\ \cline{3-4}
	      &$M$ ($\mu_B$)  & Si-rich & C-rich  \\ \hline
\multicolumn{1}{l}{Single vacancy}& &  \\ \cline{1-1}
$V_{\rm Si}$ & 4.19  & 8.053   & 9.267   \\
$V_{\rm C}$  & 1.19  & 4.419   & 3.205  \\ \\ 
\multicolumn{1}{l}{two Si vacancies} &  &  & \\ \cline{1-1}
$V_{\rm Si}^1V_{\rm Si}^2$ & -0.02 & 2.806 & 4.020 \\
$V_{\rm Si}^1V_{\rm Si}^3$ & 3.86 &  7.836 & 9.050  \\
$V_{\rm Si}^1V_{\rm Si}^4$ & 3.88 &  7.439 & 8.653 \\
$V_{\rm Si}^1V_{\rm Si}^5$ & 2.99 &  7.999 & 9.213 \\
$V_{\rm Si}^1V_{\rm Si}^6$ & 3.81 &  7.481 & 8.695 \\ \\
\multicolumn{1}{l}{Si + C vacancies} &  &  & \\\cline{1-1}
$V_{\rm Si}^1V_{\rm C}^7$ & 0.06  & 6.764 & 6.764 \\
$V_{\rm Si}^1V_{\rm C}^8$ & 1.34 & 4.682 & 4.682 \\
$V_{\rm Si}^1V_{\rm C}^9$ & 0.14 & 8.149 & 8.149 \\
$V_{\rm Si}^1V_{\rm C}^{10}$ & 1.91 &10.214&10.214\\
$V_{\rm Si}^1V_{\rm C}^{11}$ & 1.99 &11.198&11.198\\
$V_{\rm Si}^1V_{\rm C}^{12}$ & -0.93 &3.842 &3.842 \\
$V_{\rm Si}^1V_{\rm C}^{13}$ & -0.57 &4.486 & 4.486\\
$V_{\rm Si}^1V_{\rm C}^{14}$ & 4.03 &12.005 & 12.005\\
$V_{\rm Si}^1V_{\rm C}^{15}$ & 2.30 & 9.281&9.281 \\
$V_{\rm Si}^1V_{\rm C}^{16}$ & 3.47 &11.198&11.198\\
\end{tabular}
\end{ruledtabular}
\end{table}

\begin{figure}[!t]
\centering
\includegraphics[scale=0.35]{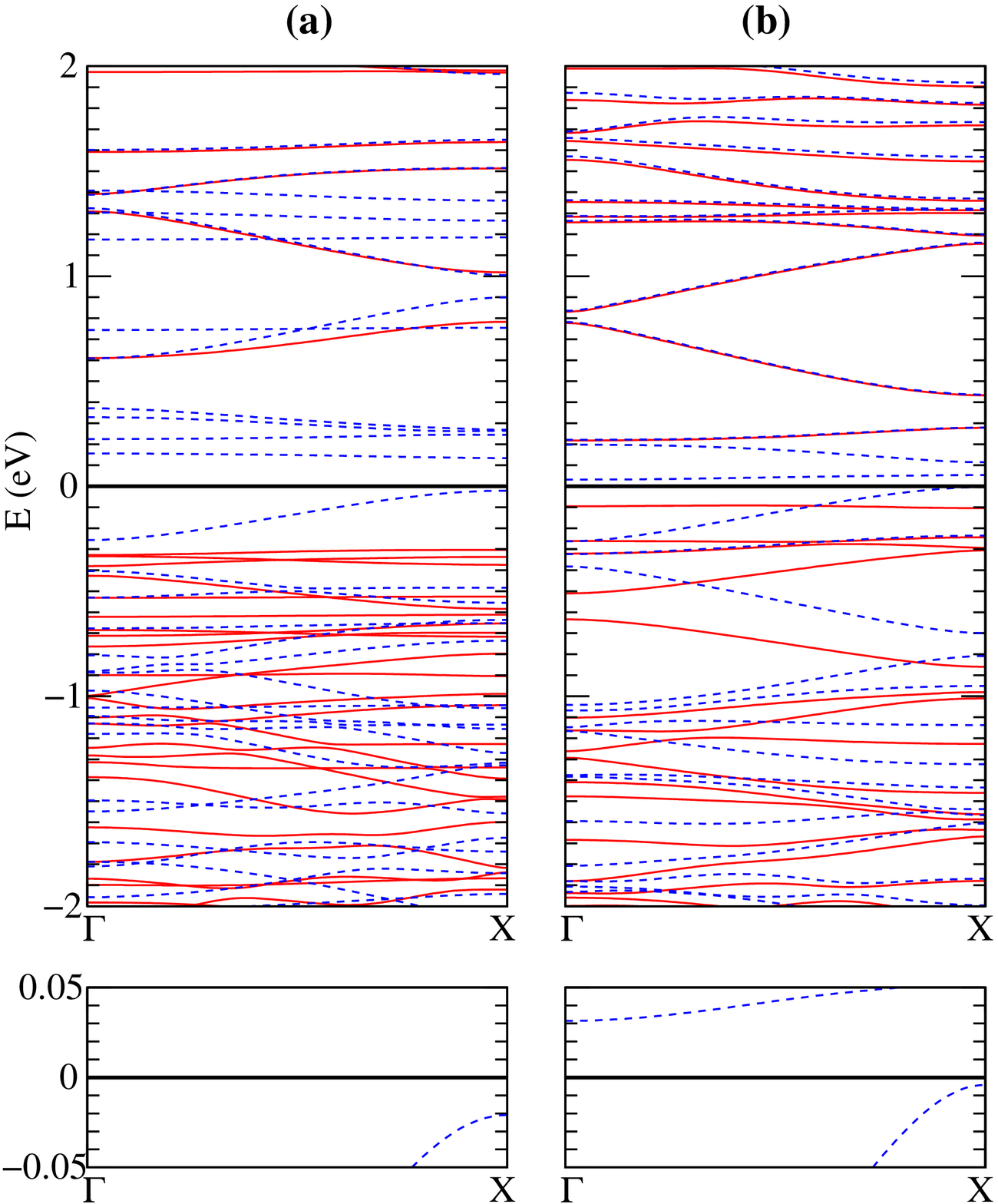}
\caption{\label{double_vacancies_bands} 
(Color online) Electronic band structure of the (a) \mbox{$V_{\rm Si}^1V_{\rm Si}^4$/6Z-SiCNR} and (b) \mbox{$V_{\rm Si}^1V_{\rm C}^{8}$/6Z-SiCNR} systems.  
The band structure of each system is presented in two ranges: 
$|E - E_F| \leq  2.0$~eV (top panel) and $|E - E_F| \leq 0.05$~eV (bottom panel). 
Solid red and dashed blue lines indicate spin-up and spin-down bands, respectively. The Fermi level is set to zero and 
indicated by the solid black line.}
\end{figure}

\begin{figure}[!t]
\centering
\includegraphics[scale=0.30]{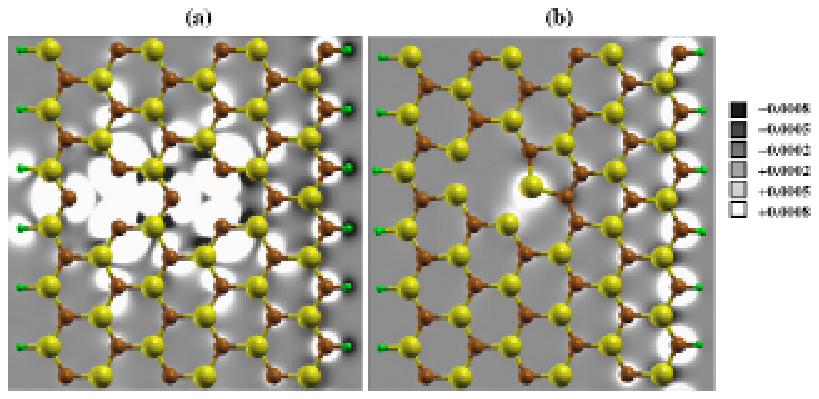}
\caption{\label{double_vacancies_spin} 
(Color online) Spin-density distribution in (a) \mbox{$V_{\rm Si}^1V_{\rm Si}^4$/6Z-SiCNR} and (b) \mbox{$V_{\rm Si}^1V_{\rm C}^{8}$/6Z-SiCNR}. 
}
\end{figure}

Figure~\ref{double_vacancies_bands} and \ref{double_vacancies_spin} present the electronic band structures [Fig.~\ref{double_vacancies_bands}]
and the spin-density distributions [Fig.~\ref{double_vacancies_spin}] 
for some representative magnetic systems with two vacancies per supercell, 
viz., \mbox{$V_{\rm Si}^1V_{\rm Si}^4$/6Z-SiCNR} and \mbox{$V_{\rm Si}^1V_{\rm C}^{8}$/6Z-SiCNR}. As can be seen in 
Fig.~\ref{double_vacancies_bands}, both \mbox{$V_{\rm Si}^1V_{\rm Si}^4$/6Z-SiCNR} and \mbox{$V_{\rm Si}^1V_{\rm C}^{8}$/6Z-SiCNR} 
systems exhibit semiconducting character, 
with band gaps of 0.91~eV  (0.15~eV) and 0.31~eV (0.04~eV) for the spin-up (-down) channels of \mbox{$V_{\rm Si}^1V_{\rm Si}^4$/6Z-SiCNR} and 
\mbox{$V_{\rm Si}^1V_{\rm C}^{8}$/6Z-SiCNR}, respectively. These results indicate that 
the presence of double vacancies ($V_{\rm Si}V_{\rm Si}$ or $V_{\rm Si}V_{\rm C}$) 
can transform half-metallic Z-SiCNRs into semiconducting systems. 
As we increased the vacancy concentrations to $\sim 5.56\%$ (in double vacancy systems), half-metal to semiconductor phase transition is observed. 
This phase transition is due to holes which are localized on the dangling bonds around the vacancies 
but also have some contribution from the edge atoms. Usually, when a vacancy is created some states around the Fermi level, 
which are occupied in the pristine system, are removed and such removal of states creates holes around the Fermi energy 
which changes the electronic structure of the material.\cite{kim,Ahn,Ana,GR}  
In Fig.~\ref{bands}(a) we see some states just below the Fermi level; however, when a Si vacancy is created, 
some missing states (near to $X$-point) can be seen [see Fig.~\ref{bands}(b)]. Similar holes were also observed in SiC monolayer with a Si vacancy.\cite{He20102451} 
In 6Z-SiCNR, the electronic and magnetic structures are not only modified by the atoms surrounding a vacancy but also 
by the edge atoms\cite{LouLee_12637} (also see Fig.~4(d)). This makes 6Z-SiCNR different from other materials where magnetism is caused by vacancies.
Our results teach us that the electronic band structure of Z-SiCNRs can easily be engineered by 
$V_{\rm Si}$ or $V_{\rm C}$. 
The distribution of spin density in \mbox{$V_{\rm Si}^1V_{\rm Si}^4$/6Z-SiCNR} [Fig.~\ref{double_vacancies_spin}(a)] shows that 
the magnetic moments are mostly localized on the C atoms surrounding the Si vacancies. 
The C atoms around $V_{\rm Si}^4$ have magnetic moments of 0.23, 0.88 and $0.88\mu_B$, while the C atoms around $V_{\rm Si}^1$ 
have magnetic moments of 1.06, 0.91 and $0.91\mu_B$. 
On the other hand, in \mbox{$V_{\rm Si}^1V_{\rm C}^{8}$/6Z-SiCNR} [Fig.~\ref{double_vacancies_spin}(b)] 
we observe that the C atoms surrounding $V_{\rm Si}^1$ move to form new C-C bonds; thus, the dangling bonds of these atoms are recombined 
leading to almost zero local magnetic moments. In this system the magnetic moments are mostly localized on edge C atoms: we found magnetic moments 
between 0.16 and $0.21\mu_B$ for each edge C atom.

As we can see, in the $V_{\rm Si}^1V_{\rm Si}^4$ system one of the Si vacancies lies close to the edge [see Fig.~\ref{double_vacancies_spin}(a)]. 
To ignore the edge effect (if it has some effect on the electronic and magnetic properties of SiCNRs) we also carried out some test calculations 
for $V_{\rm Si}^1V_{\rm Si}^4$ by considering a large supercell (12Z-SiCNR). In this case, we kept the same distance between $V_{\rm Si}^1$ 
and $V_{\rm Si}^4$ that we used in the small supercell. The calculated magnetic moment (per supercell and per Si vacancy) is $\sim 3.50\mu_B$. 
{We must note that the small difference in the magnetic moments (3.88$\mu_B$ in 6Z-SiCNR and 3.50$\mu_B$ in 12Z-SiCNR) 
is due to the difference in the total number of atoms in the two NRs.}

As our studied systems induce magnetism due to vacancies, and the magnetic moments are not carried by conventional magnetic elements (Fe, Ni, or Co) 
where one can also calculate the true ground magnetic state of a material. Either $V_{\rm Si}$ or $V_{\rm C}$ in 6Z-SiCNR distorts the bond lengths 
which can give different atomic magnetic moments at different atomic sites. To get a qualitative analysis of magnetism in 6Z-SiCNR, we also carried out calculations 
by considering the antiferromagnetic (AFM) and  ferromagnetic (FM) types of $V_{\rm Si}-V_{\rm Si}$ interactions. 
{We studied the FM and AFM interactions between the Si vacancies by considering the FM and AFM coupling between the atoms 
surrounding the vacancies. We solved the Kohn-Sham equation~\cite{Kohn} in the FM and AFM states and compared their total energies 
to get the ground state magnetic structure.} 
We found that the FM  state is more 
stable than the AFM state by $\sim 0.14$ eV per cell which is much larger than room temperature. Therefore, room-temperature magnetism is expected in 6Z-SiCNR. 
The FM stability against AFM was also confirmed through fixed moment calculations.
It is emphasized that such kind of magnetic coupling strongly depends on the distance between the vacancies.\cite{Sergey,GR} 
{Some of the  selected cases were also reinvestigated using a large supercell to consider the FM/AFM coupling between the Si vacancies. 
We noticed that FM is still expected. }

\subsection{Interaction of vacancies with substitutional B and N impurities}

Among the defects investigated, single $V_{\rm Si}$ was found to induce the largest magnetic moment [see Table~\ref{table1}], 
which suggests that this defect can play an important role on the magnetism of Z-SiCNR. 
However, the formation of single $V_{\rm Si}$ in Z-SiCNR is expected to be energetically unfavorable due to its high defect formation energy. 
To explore this system, we need to reduce its defect formation energy and for this reason the role of substitutional B and N impurities in  
6Z-SiCNR is also investigated. 
Since Costa and Morbec\cite{Costa_Morbec_2011} recently reported that B prefers to occupy a Si site whereas N preferentially substitutes a C atom,
we examined configurations with single $V_{\rm Si}$ at position 1 and (i) B at Si site $i=2,...,6$ [see Fig.~\ref{supercell}], 
\mbox{$V_{\rm Si}^1B_{\rm Si}^i$/6Z-SiCNR}, or (ii) N at C site $i=7,...,16$, 
\mbox{$V_{\rm Si}^1N_{\rm C}^i$/6Z-SiCNR}. 
The energetic stability of these systems was determined from their formation energies 
\begin{equation}
E_f=E[\mbox{$V_{\rm Si}^1Y_{\rm X}^i$/6Z-SiCNR}]-E[\mbox{6Z-SiCNR}]+\mu_{\rm Si}+\mu_{\rm X}-\mu_{\rm Y},
\end{equation}
where X=Si and Y=B for the \mbox{$V_{\rm Si}^1B_{\rm Si}^i$/6Z-SiCNR} systems, and 
X=C and Y=N for \mbox{$V_{\rm Si}^1N_{\rm C}^i$/6Z-SiCNR}. 
We used $\alpha$-Boron bulk and N$_2$ molecule to obtain B and N chemical potentials, respectively.

\begin{table}[!t]
\caption{\label{table2} Calculated formation energies, in Si-rich and C-rich conditions, and total magnetic moments per supercell ($M$)  
for 6Z-SiCNR with a single $V_{\rm Si}$ and B (N) at Si (C) site. 
The index $i$ in $V_{\rm Si}^1B_{\rm Si}^i$ and $V_{\rm Si}^1N_{\rm C}^i$ 
indicates the substitutional Si or C site for B or N impurity [see Fig.~\ref{supercell}]. }
\begin{ruledtabular}
\begin{tabular}{lccc}
              &  & \multicolumn{2}{c}{Formation energies (eV)} \\ \cline{3-4}
	      &$M$ ($\mu_B$)  & Si-rich & C-rich  \\ \hline
\multicolumn{1}{l}{B at Si site} &  &  & \\ \cline{1-1}
$V_{\rm Si}^1B_{\rm Si}^2$ &  2.86 & 6.904 & 9.332 \\
$V_{\rm Si}^1B_{\rm Si}^3$ & 4.90 &  8.235 & 10.663 \\
$V_{\rm Si}^1B_{\rm Si}^4$ & 2.24 &  7.058 &  9.486\\
$V_{\rm Si}^1B_{\rm Si}^5$ & 4.61 &  8.611 & 11.039\\
$V_{\rm Si}^1B_{\rm Si}^6$ & 4.81 &  7.565 &  9.993\\ \\
\multicolumn{1}{l}{N at C site} &  &  & \\\cline{1-1}
$V_{\rm Si}^1N_{\rm C}^7$ & 0.63  & 5.133 & 5.133 \\
$V_{\rm Si}^1N_{\rm C}^8$ & 3.06 & 8.246 & 8.246 \\
$V_{\rm Si}^1N_{\rm C}^9$ & 3.07 & 7.662 & 7.662 \\
$V_{\rm Si}^1N_{\rm C}^{10}$ & 3.01 &8.583 &8.583 \\
$V_{\rm Si}^1N_{\rm C}^{11}$ & 2.38 &7.188 & 7.188\\
$V_{\rm Si}^1N_{\rm C}^{12}$ &  3.12 &7.417 &7.417 \\
$V_{\rm Si}^1N_{\rm C}^{13}$ &  3.12 &7.682 & 7.682\\
$V_{\rm Si}^1N_{\rm C}^{14}$ & 3.09 & 7.660 &  7.660\\
$V_{\rm Si}^1N_{\rm C}^{15}$ & 3.04 & 8.212&8.212 \\
$V_{\rm Si}^1N_{\rm C}^{16}$ & 2.95 &8.956 &8.956 \\
\end{tabular}
\end{ruledtabular}
\end{table}

The formation energies (in Si-rich and C-rich conditions) and the total magnetic moments per supercell 
for \mbox{$V_{\rm Si}^1B_{\rm Si}^i$/6Z-SiCNR} and \mbox{$V_{\rm Si}^1N_{\rm C}^i$/6Z-SiCNR} systems are listed 
in Table~\ref{table2}.  
{From these results, we note that some \mbox{$V_{\rm Si}^1B_{\rm Si}^i$/6Z-SiCNR} and  
\mbox{$V_{\rm Si}^1N_{\rm C}^i$/6Z-SiCNR} configurations are energetically more favorable than \mbox{$V_{\rm Si}^1$/6Z-SiCNR} (for example, 
$V_{\rm Si}^1B_{\rm Si}^2$ and $V_{\rm Si}^1B_{\rm Si}^4$ are more favorable than \mbox{$V_{\rm Si}^1$/6Z-SiCNR} at Si-rich and stoichiometric conditions). 
This indicates that one B atom occupying a Si site, as well as one N atom occupying a C site, can stabilize 
a single $V_{\rm Si}$ in Z-SiCNR.}
In addition, we observe that the \mbox{$V_{\rm Si}^1B_{\rm Si}^i$/6Z-SiCNR} and 
\mbox{$V_{\rm Si}^1N_{\rm C}^i$/6Z-SiCNR} systems have large magnetic moments 
(except for \mbox{$V_{\rm Si}^1N_{\rm C}^7$/6Z-SiCNR} whose magnetic moment is $0.63\mu_B$). 
This smallest magnetic moment can be expected in \mbox{$V_{\rm Si}^1N_{\rm C}^7$/6Z-SiCNR} based on the separation between 
\mbox{$V_{\rm Si}^1-N_{\rm C}^7$} which is 1.81\AA. 
We found magnetic moments larger than $2\mu_B$ for all other configurations with $V_{\rm Si}B_{\rm Si}$ or $V_{\rm Si}N_{\rm C}$. 
However, the \mbox{$V_{\rm Si}N_{\rm C}$/6Z-SiCNR} systems are more interesting than \mbox{$V_{\rm Si}B_{\rm Si}$/6Z-SiCNR}. 
When N is doped at the vacant C site in \mbox{$V_{\rm Si}V_{\rm C}$/6Z-SiCNR}, then the C-\mbox{$V_{\rm Si}$}-types interactions increase the magnetization. 
So, we believe that N in 6Z-SiCNR has twofold roles; it not only stabilizes the intrinsic defects but also increases the magnetization of 6Z-SiCNR.  
From the whole thermodynamic data, it is summarized that those systems with low magnetic moments have smaller formation energies.

\begin{figure}[!t]
\centering
\includegraphics[scale=0.35]{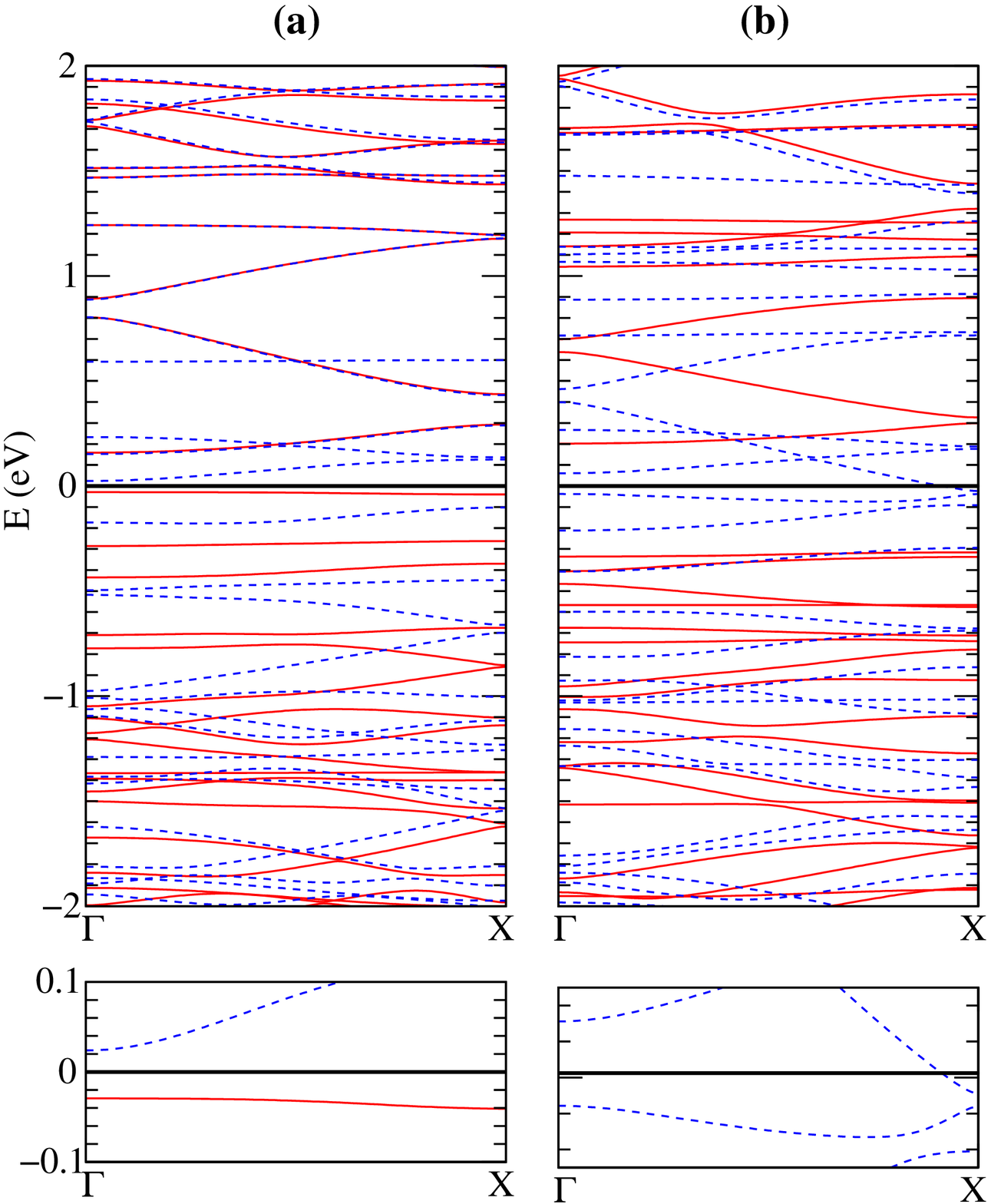}
\caption{\label{impurity_bands} 
(Color online) Electronic band structure of the (a)~\mbox{$V_{\rm Si}^1B_{\rm Si}^4$/6Z-SiCNR} and (b) \mbox{$V_{\rm Si}^1N_{\rm C}^{8}$/6Z-SiCNR} systems.  
The band structure of each system is presented in two ranges: 
$|E - E_F| \leq  2.0$~eV (top panel) and $|E - E_F| \leq 0.1$~eV (bottom panel). 
Solid red and dashed blue lines indicate spin-up and spin-down bands, respectively. The Fermi level is set to zero and 
indicated by the solid black line.}
\end{figure}

The electronic band structures for the \mbox{$V_{\rm Si}^1B_{\rm Si}^4$/6Z-SiCNR} and \mbox{$V_{\rm Si}^1N_{\rm C}^{8}$/6Z-SiCNR} systems are 
depicted in Fig.~\ref{impurity_bands}. As can be seen, the semiconducting character of \mbox{$V_{\rm Si}^1V_{\rm Si}^4$/6Z-SiCNR} 
is maintained when B is doped at the vacant Si$^4$ site:  both the spin-up and spin-down channels of \mbox{$V_{\rm Si}^1B_{\rm Si}^4$/6Z-SiCNR} 
[Fig.~\ref{impurity_bands}(a)] are semiconducting with band gaps of 0.19 and 0.13~eV, respectively. 
However, the presence of a N atom occupying a vacant C site can lead semiconducting \mbox{$V_{\rm Si}V_{\rm C}$/6Z-SiCNR} systems to 
turn into half-metallic ones.  Fig.~\ref{impurity_bands}(b) shows that the \mbox{$V_{\rm Si}^1N_{\rm C}^{8}$/6Z-SiCNR} system is a half-metal: while its 
spin-down channel is metallic, the spin-up channel is semiconducting with band gap of 0.52~eV. 
It is interesting to note that \mbox{$V_{\rm Si}^1B_{\rm Si}^4$/6Z-SiCNR} remains semiconducting consisting with our picture of hole-induced electronic 
phase transition. When a vacant $V_\mathrm{Si}^{i}$ site is filled with boron in bivacancy $V_\mathrm{Si}^{1}V_\mathrm{Si}^{i}$ system, it means that we 
are filling the empty orbitals (holes) created by $V_\mathrm{Si}^{i}$. However, the \mbox{$V_{\rm Si}^1B_{\rm Si}^4$} system has sufficient holes, 
compared with the pristine  6Z-SiCNRs, and remains  semiconducting as expected. Boron mainly induces impurity states near the Fermi level in both spin-up 
and spin-down channels [see Fig.~\ref{impurity_bands}(a)]. On the other hand when N is doped at $V_\mathrm{C}$ site in $V_\mathrm{Si}^{1}V_\mathrm{C}^{i}$ 
system, electrons are injected into $V_\mathrm{Si}^{1}V_\mathrm{C}^{i}$ and we have sufficient electrons to fill the empty states. 
These electrons are mostly in the spin-down states and have sufficient energy to cross the Fermi energy, and $V_\mathrm{Si}^{1}N_\mathrm{C}^{8}$, e.g., 
transforms to half-metal as compared with $V_\mathrm{Si}^{1}V_\mathrm{C}^{8}$. Such re-entrant behavior of 6Z-SiCNR is ascribed to electron doping.

\begin{figure}
\centering
\includegraphics[scale=0.30]{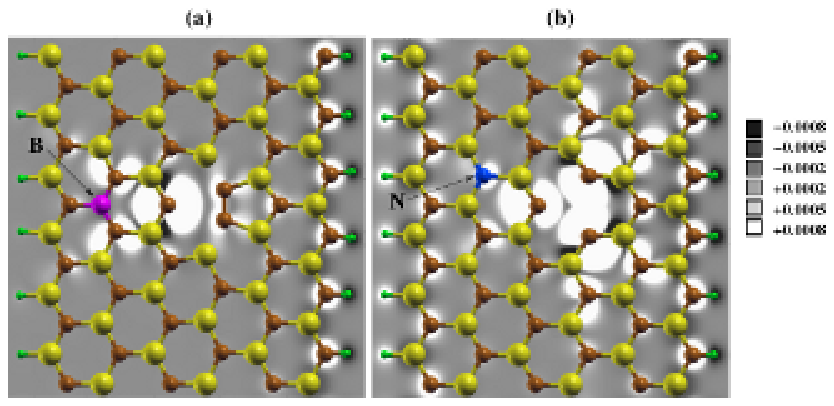}
\caption{\label{impurity_spin} 
(Color online) Spin-density distribution in (a)~\mbox{$V_{\rm Si}^1B_{\rm Si}^4$/6Z-SiCNR} and (b) \mbox{$V_{\rm Si}^1N_{\rm C}^{8}$/6Z-SiCNR}. 
}
\end{figure}

The spin-density maps presented in Fig.~\ref{impurity_spin} show that the presence of a B (N) impurity filling a vacant 
Si (C) site strongly affects the distribution of spin-density in \mbox{$V_{\rm Si}V_{\rm Si}$/6Z-SiCNR} (\mbox{$V_{\rm Si}V_{\rm C}$/6Z-SiCNR}). 
When a B atom occupies the vacant ${\rm Si}^4$ site [see Fig.~\ref{impurity_spin}(a)] 
the dangling bonds of the C atoms surrounding $V_{\rm Si}^4$ are saturated leading the magnetic moments of these C atoms to decrease 
from 0.23, 0.88 and $0.88\mu_B$ in \mbox{$V_{\rm Si}^1V_{\rm Si}^4$/6Z-SiCNR} to 0.004, 0.039 and $0.039\mu_B$ in \mbox{$V_{\rm Si}^1B_{\rm Si}^4$/6Z-SiCNR}. 
In addition, two of the three C atoms surrounding $V_{\rm Si}^1$ move closer to each other to form a C-C bond, which leads to a decrease 
(from $0.91\mu_B$ in \mbox{$V_{\rm Si}^1V_{\rm Si}^4$/6Z-SiCNR} to $0.02\mu_B$ in \mbox{$V_{\rm Si}^1B_{\rm Si}^4$/6Z-SiCNR}) 
in the magnetic moments of these atoms; the third C atom surrounding $V_{\rm Si}^1$ (labeled 7 in Fig.~\ref{supercell}) has magnetic moment of about $0.80\mu_B$. 
On the other hand, Fig.~\ref{impurity_spin}(b) shows that bond reconstructions in the vicinity of $V_{\rm Si}^1$, which were observed 
in \mbox{$V_{\rm Si}^1V_{\rm C}^{8}$/6Z-SiCNR}, do not occur when N is doped at the vacant C$^8$ site. 
Such a process results in a increase in the local magnetic moments of the C atoms surrounding $V_{\rm Si}^1$ 
(we found magnetic moments of 0.17, 0.87 and $0.90\mu_B$ in these atoms) and, 
consequently, in a increase in the magnetization of the system (\mbox{$V_{\rm Si}^1V_{\rm C}^{8}$/6Z-SiCNR} has magnetic moment of $1.34\mu_B$, while 
\mbox{$V_{\rm Si}^1N_{\rm C}^{8}$/6Z-SiCNR} has magnetic moment of $3.06\mu_B$).

{Before summarizing our DFT calculations, we would like to comment on the shortcomings of DFT-PBE calculations. 
This is an accepted fact in the DFT community that DFT-LDA/GGA calculations do not reproduce the experimental band gap of insulators and semiconductors. 
The band gap  of SiC can be recovered either using LDA/GGA+U or GW or SIC type calculations.\cite{prb-1995,prb-2002,prb-2006} 
However, there are DFT+U calculations which clearly demonstrate that defect(vacancy)-induced magnetism does not change either using DFT or DFT+U 
calculations.~\cite{Sanvito,Fernandes,gr2013} Indeed DFT+U calculations will affect the electronic structure of materials. 
We believe that DFT+U/SIC/GW calculations will not change the  magnetism of SiC NRs.}
 
\section{CONCLUSIONS} \label{conclusion}

{\it Ab-initio} calculations were performed to investigate the magnetism of pristine and defected Z-SiCNRs. 
Single ($V_{\rm Si}$ and $V_{\rm C}$) and double ($V_{\rm Si}V_{\rm Si}$ and $V_{\rm Si}V_{\rm C}$) vacancies were considered to probe the electronic 
structures of Z-SiCNRs. Our results indicate that these native defects can induce large magnetic moments in Z-SiCNRs. 
While the half-metallic character of the pristine Z-SiCNR is maintained in the presence of a single $V_{\rm Si}$, 
we found that a single $V_{\rm C}$ leads to a transition from half-metallic to metallic behavior in Z-SiCNRs, and 
double $V_{\rm Si}V_{\rm Si}$ and $V_{\rm Si}V_{\rm C}$ can transform half-metallic Z-SiCNRs into semiconducting systems. 
Such electronic phase transitions were discussed in terms of hole doping. 
Single $V_{\rm Si}$ induced the largest magnetic moment among the defects investigated, 
but this defect has higher formation energy and is energetically unfavorable when compared with the single $V_{\rm C}$ and double $V_{\rm Si}V_{\rm C}$.
The ferromagnetic ground state was shown to be more stable than the  antiferromagnetic state; 
therefore, room-temperature ferromagnetism was also speculated in defected Z-SiCNRs. 
To reduce the defect formation energy of single $V_{\rm Si}$ and realize it experimentally, the interactions of substitutional B and N impurities 
with this native defect were also studied. 
We found that a B atom substituting a Si atom, as well as a N atom substituting a C atom in the presence of vacancies, 
leads to a considerable reduction of the defect formation energy of $V_{\rm Si}$. 
Therefore, we believe that light elements are beneficial for the realization of room-temperature magnetism in defective Z-SiCNRs.

\section*{ACKNOWLEDGMENTS}

We are grateful to V\'{\i}ctor M. Garc\'{\i}a-Su\'arez for useful discussions. 
JMM acknowledges computational support from CENAPAD/SP (Brazil). GR acknowledges the cluster facilities of NCP, Pakistan.


\begin{thebibliography}{25}
\bibitem{footnote}{Our Mulliken population analysis indicates 
that when a C vacancy is created, the total charge on the Si atoms surrounding the C vacancy is decreased,  
and the holes reside on both spin-up and spin-down channels of the Si atoms, which can be seen in Figs.~\ref{bands}(c) and \ref{density_states}(c).}
\end{thebibliography}
\end{document}